\documentclass[10pt]{iopart}

\usepackage{amssymb}
\usepackage{graphicx}
\usepackage{epstopdf}
\usepackage{textcomp}
\usepackage{color}
\usepackage{filecontents}
\usepackage{cite}
\usepackage[square,sort,comma,numbers]{natbib}
\usepackage{url}
\usepackage{iopams} 
\expandafter\let\csname equation*\endcsname\relax
\expandafter\let\csname endequation*\endcsname\relax 
\usepackage{amsmath}
\usepackage{amsfonts} 

\begin{document}

\title[Confined H$_2^+$]{Spherically confined H$_2^+$: $^2\Sigma_g^+$ and $^2\Sigma_u^+$ states}
\author{G. Micca Longo$^1$ , S. Longo$^{1,2,3}$ and D. Giordano$^4$}

\address{$^1$ CNR-IMIP, via Amendola 122/D, 70126 Bari, Italy}
\address{$^2$ Department of Chemistry, University of Bari, via Orabona 4, 70126 Bari, Italy}
\address{$^3$ INAF-Osservatorio Astrofisico di Arcetri, Largo E. Fermi 5,
              I-50125 Firenze, Italy}
\address{$^4$ ESA-ESTEC, Aerothermodynamics Section, Kepleerlaan 1, 2200 ag, Noordwijk,
              The Netherlands}

\ead{gaia.miccalongo@imip.cnr.it}
\ead{savino.longo@imip.cnr.it}
\ead{domenico.giordano@esa.int}

\vspace{10pt}
\begin{indented}
\item[]February 2014
\end{indented}

\begin{abstract}
The molecular ion H$_2^+$ is studied under strong confinement conditions produced by a spherical barrier centered in the gravity center of the molecule. Results for the potential curves are obtained by diffusion Monte Carlo methods for the ground state (X$^2\Sigma_g^+$) and the first excited state (A$^2\Sigma_u^+$), and reported as functions of the internuclear distance d for different values of the confinement radius. Results show that the compressed state corresponding to both $^2\Sigma_g^+$ and $^2\Sigma_u^+$ present deep minima in their potential curves, due to the increased space for electron wave-functions when the protons are displaced from the barrier surface. 
\end{abstract}

\noindent{\it Keywords}: Confined H$_2^+$, DMC, High pressures

\ioptwocol

\section{Introduction}
\label{intro}
Simple confined physical systems have a very long tradition in physics 
\citep{michels1937,degroot1946, sury1976,yngve1988,yngve1986,burrows2006,aquino2007},
because of their importance as basic theoretical problems and as models for microscopic components of compressed matter in 
astrophysics (interior of stars), geophysics (giant planets), atomic and semiconductor physics (quantum dots).

In particular, confined H$_2^+$, the simplest molecular system, has been the subject of many investigations as a paradigmatic model for the analysis of molecular systems under pressure: different approches have been applied to the model of H$_2^+$ confinement \citep{cottrell1951,singh1964,leykoo1981,lesar1981,gorecki1988,
mateoscortes2002,skouteris2010,colin2011,sarsa2012}.

In most previous papers, ion positions are bound to the confined surface geometry
by fixed geometrical relations.  Furthermore, most studies assume prolate ellipsoidal confinement: this condition is suggested by the fact that, as known, the Schr\"{o}dinger equation for H$_2^+$ factorizes using prolate ellipsoidal coordinates. Spherical confinement where considered very sparingly \citep{gorecki1988,burrows2013} using rather analytically heavy variational approaches. 
It is clear that a more straightforward approach is desirable also in view of the interest of the subject.

In this paper we apply the diffusion Monte Carlo (DMC) method to the problem of spherically confined H$_2^+$.
DMC is a very versatile method for such applications since an accurate value of energy can be determined using cartesian coordinates with a very natural coexistence of the different symmetries of the the two-nucleus coulomb field and the spherical barrier. Details of the method can be found in a huge literature  \citep{foulkes2001,reynolds1982,umrigar1993}. In
the case of H$_2^+$, DMC method provides an exact solution for the
spherical confinement (unlike spheroidal like in previous studies),
moving freely the ions inside it and strongly
reducing the sphere radius, thereby raising the pressure to very high
values. These
possibilities are illustrated in the paper. A spherical geometry has been chosen because it is simpler and it is able to  show the strong points of the DMC method. Both the ground state and
the first excited state of H$_2^+$ are considered. Additionally, the use of
theoretical concepts inspired by the DMC methodology allows to estimate
some quantities and rationalize results.

We show that, contrary to intuition, H$_2^+$ is a stable
system even into very small confining spheres. Actually, the
lowest energy electronic state is a bound state with the same
qualitative behavior it has in the vacuum. Even more remarkably, the repulsive excited state in the case of free ion becomes a bound one when the ion is compressed. This behavior is
explained by the compression of the electronic wave function by the
confining potential well.

\section{Diffusion Monte Carlo}
\label{DMC}
Quantum Monte Carlo (QMC) is a class of computer algorithms that are able to simulate quantum systems and to compute the electronic ground state of atoms, molecules and solids.

Diffusion Monte Carlo (DMC), in particular, is a stochastic projector method that makes use of the similarity between the imaginary-time Schr\"{o}dinger equation and a generalized diffusion equation, which can be solved using a stochastic calculus and simulating a random walk. 
A computational code has been developed by the authors and validated as shown in the next section; no importance sampling transformation has been performed in the present paper.  

When the DMC method is applied to this problem, the hamiltonian  can be expressed in cartesian coordinates and there is no advantage to consider prolate spheroidal coordinates. Therefore, any confinement geometry is straightforward implemented, in particular a spherical confinement like here, and any position of nuclei is equivalent. Based on these features, in perspective, a comparison between different confinement conditions is possible, in particular to compare ellipsoidal \citep{burrows2013} and spherical barriers.

In our simulations, the dihydrogen cation H$_2^+$ is placed inside an impenetrable spherical box. The two protons are considered as point sources of a Coulomb field and they are placed along the x axis at $x=\frac{d}{2}$ and $x=-\frac{d}{2}$ respectively ($d$ being the intenuclear distance). The corresponding hamiltonian (in atomic units) is:

\begin{equation}\label{1}
\hat{H}=-\frac{1}{2}\nabla^2+V(\textbf{r})+\frac{1}{d}+E_H
\end{equation}

where $E_H$ is the hydrogen atom energy and $V(\textbf{r})$ is the potential energy

\begin{equation}\label{2}
V(\textbf{r})=-\frac{1}{|\textbf{r}-\textbf{i}\frac{d}{2}|}-\frac{1}{|\textbf{r}+\textbf{i}\frac{d}{2}|}
\end{equation}

 The confining barrier is a sphere of radius $r_0$ with center in the origin of the axes.
The electron is replaced by a chain of fictitious particles called walkers \citep{foulkes2001} and the Coulomb potential acts on every walker. Any walker for which $r>r_0$ ($r_0$ being the confinement dimension) is removed from the simulation in order to include the confinement. The confinement is realised by acting on the potential energy surface $V$: $V=\infty$ when $r>r_0$. By decreasing the
timestep and increasing the walker number, this method, as known, converges to the exact solution of the Schr\"{o}dinger equation.

During the simulation, the walkers diffuse throughout the phase space and the transition probability density for the evolution of the walkers is given by the approximate Green's function:

\begin{equation}\begin{split}\label{1}
G(\textbf{R}\leftarrow\textbf{R'},\tau)\approx \left(2\pi\tau\right)^{-\frac{3N}{2}}exp\left[-\frac{\left(\textbf{R}-\textbf{R'}\right)^2}{2\tau}\right]\\
exp\left[-\frac{\tau\left(V\left(\textbf{R}\right)-V\left(\textbf{R'}\right)-2E_T\right)}{2}\right]
\end{split}\end{equation}

Every diffusion step consists of two phases: propagation and branching. At the beginning, 
each walker is moved from its old position \textbf{R} to the new one \textbf{R'} with probability

\begin{equation}\label{1}
T=\left(2\pi\tau\right)^{-\frac{3N}{2}}exp\left[-\frac{\left(\textbf{R}-\textbf{R'}\right)^2}{2\tau}\right]
\end{equation}
where $\tau$ is the time step.

The factor that determines the number of walkers surviving for next step is given by

\begin{equation}\label{2}
P=exp\left[-\frac{\tau\left(V\left(\textbf{R}\right)-V\left(\textbf{R'}\right)-2E_T\right)}{2}\right]
\end{equation}
in which $V$ is the potential energy and $E_T$ is the so called energy offset that controls the total population of the walkers. When $P<1$, the walker continues its evolution with 
probability $P$ and dies with probability $1-P$; when $P\geq1$, the walker continues its evolution and, at the same position, a new walker is created with probability $P-1$. From equation (\ref{2}), it is clear that the walkers tend to proliferate in regions of low potential energy and to disappear in regions of high potential energy.

The energy offset $E_T$ is determined by keeping track of changing walkers and by tuning it at every step in order to make the average walker population approximately constant \cite{thijssen}. A simple formula for adjusting $E_T$ is

\begin{equation}\label{3}
E_{T_i}=E_{T_{i-1}}+\alpha\ln\left(\frac{N_{i-1}}{N_i}\right)
\end{equation}
where $E_{T_{i-1}}$ is the energy value at time step $i-1$, $\alpha$ is a small positive parameter and ${N_{i-1}}$ and $N_i$ are respectively the number of walkers at time step $i-1$ and the actual one. 

Probabilistic method like DMC can deal with positive distributions only, so a problem arises studying an excited state wave function: the DMC algorithm is not able to maintain the fermionic symmetry of the excited state and so the solutions fall on the bosonic ground state.
Nevertheless, DMC method can still be used for low excited states with distinct symmetries. The starting point is the fixed-node DMC method \citep{reynolds1982, anderson1975,anderson1976,moskowitz1982}: a trial wave function is chosen and used to define a trial node surface that reproduces 
the symmetry of the excited state.

Here the $^2\Sigma_u^+$ eigenvalues are reproduced by an extra absorbing boundary placed at the centre of the box, corresponding to the $yz$ plane.

A problem arising during computation and associated with the confinement boundary requires attention: the boundary is not truly \textit{numerically} impenetrable
and so it is possible that some walker does not see the wall and passes through it. 
As a matter of fact, the ensemble of walkers behaves like a kind of rarefied gas with a mean free path $\sim\sqrt{\tau}$; therefore, if the estimator is sampled after checking 
the barrier crossing, the effective box width is larger by a quantity of the same order.
An optimization technique, based on a sub-cycling algorithm, can be applied: walkers getting closer than a few new $\sqrt{\tau}$ 
to the barrier can be moved with a reduced step, typically $0.1\tau$, for a number of cycles chosen to match the original $\tau$. This was not found necessary to provide the plot in this work.

The DMC method can also be applied to partial confinement: outside the confining sphere, the potential energy $V$ can be set to a large, but not infinity, value, in order to realise a penetrable confinement \citep{colin2011}.

These last two observations provide an interesting starting point for future works.

\section{Confined H$_2^+$}
\label{conf}

Here results for the H$_2^+$ electronic ground state and the first excited state are reported. About 10$^3$ walkers have been used in the calculation, while $\tau$ values ranged between 10$^{-3}$ for strong confinement and the excited state and 10$^{-2}$ for free case. The total energy (presented in $eV$) as a function of the internuclear separation $d$ (presented in atomic units) is determined.
 
Preliminary, the calculations are validated by comparing the results for the two states with results from the literature, for the case of an unconfined ion, since no result is available for the confined system. 

\begin{figure}
\resizebox{0.5\textwidth}{!}{
  \includegraphics{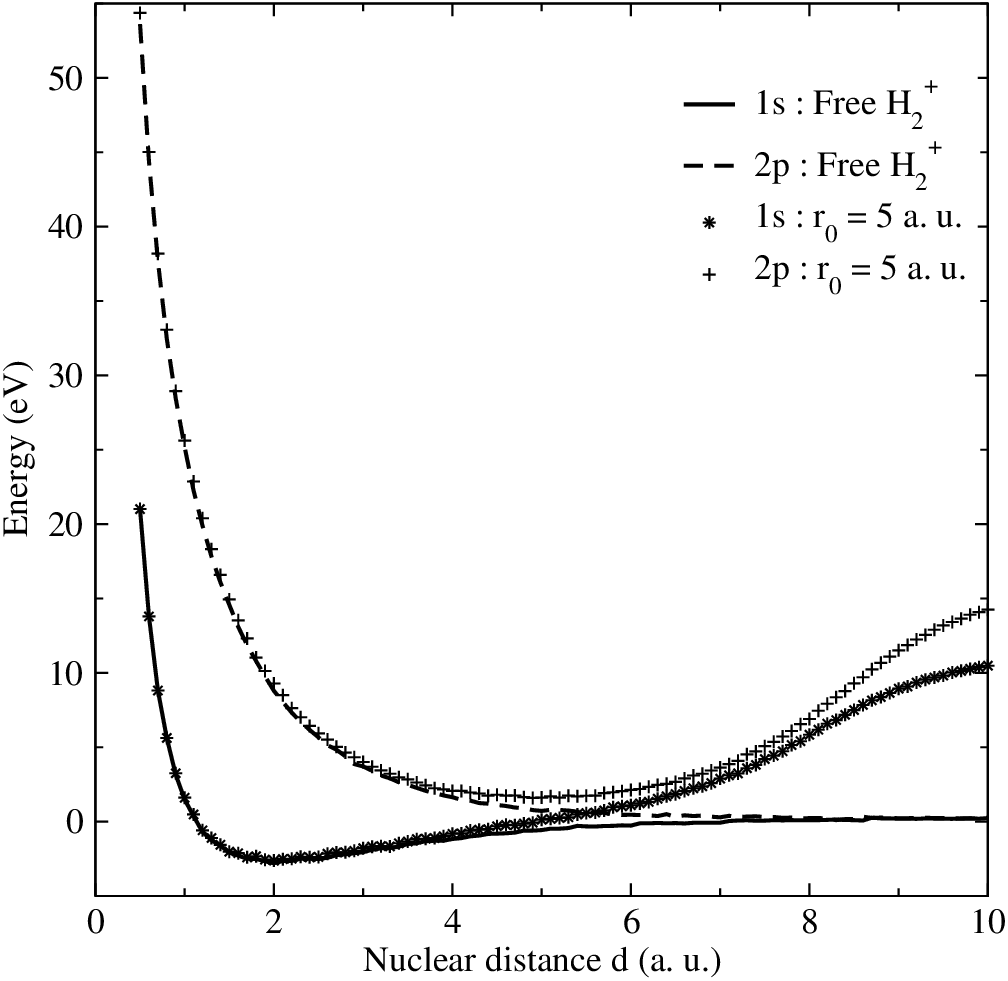}
}
\caption{Free and confined H$_2^+$.}
\label{fig:1} 
\end{figure}

In Figure \ref{fig:1} energy eigenvalues are reported as a function of $d$, for the two different states of the free and mildly compressed H$_2^+$. This plot is reproduced here for simpler comparison to confined states, while at the same time it provides a validation of the code used for calculations: at the scale of the plot, the curves for the free case are indistinguishable from the state of the art results.

\begin{figure}
\resizebox{0.5\textwidth}{!}{
  \includegraphics{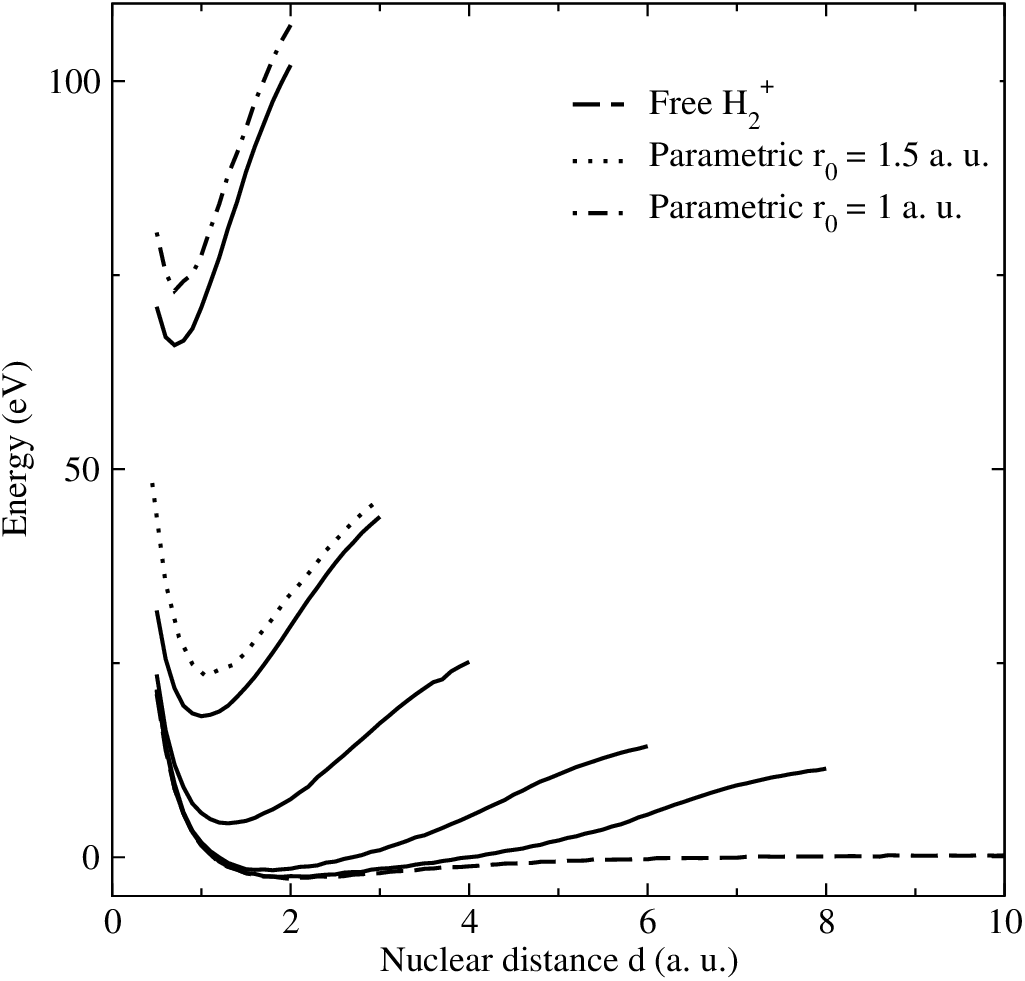}
}
\caption{Free and confined H$_2^+$: ground state. For better detection, note that for each $r_0$ the corresponding continue curve ends at $d=2 r_0$.}
\label{fig:2} 
\end{figure}

\begin{figure}
\resizebox{0.5\textwidth}{!}{
  \includegraphics{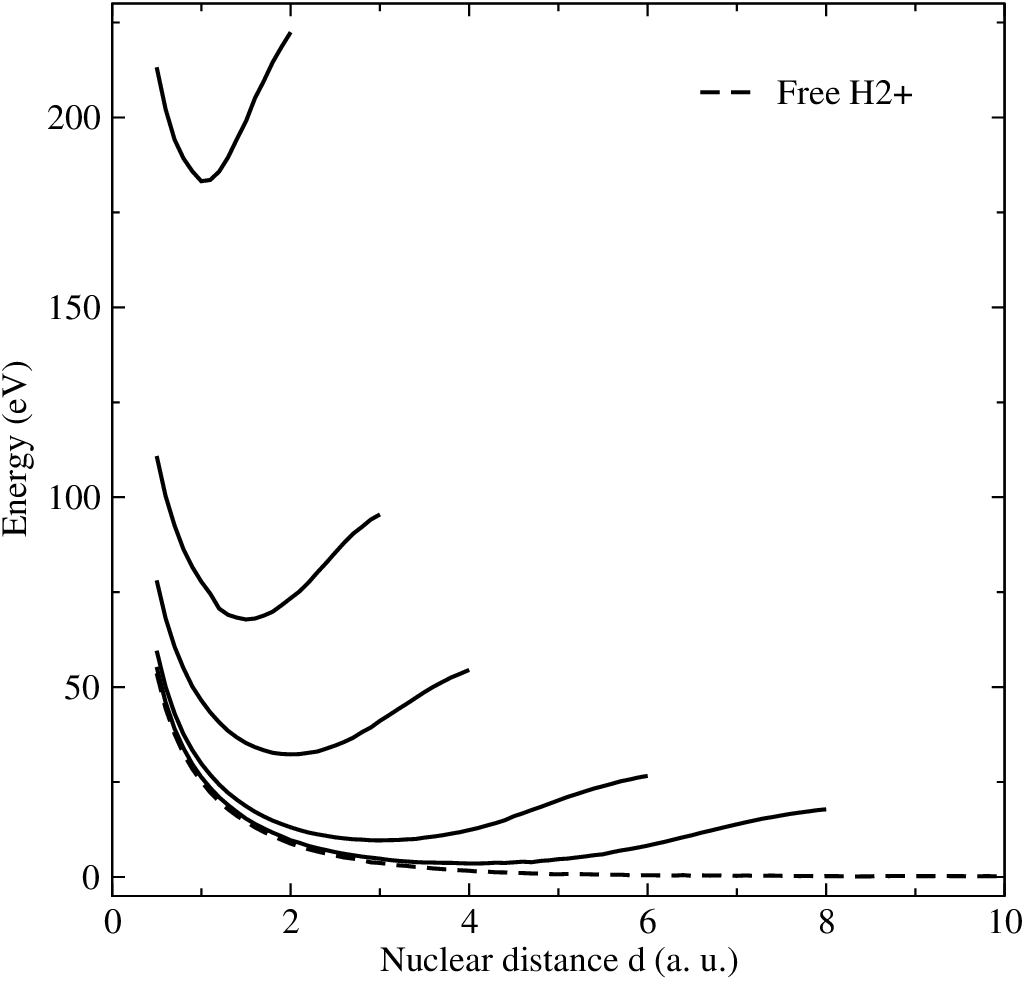}
}
\caption{Free and confined H$_2^+$: first excited state. For better detection, note that for each $r_0$ the corresponding continue curve ends at $d=2 r_0$.}
\label{fig:3} 
\end{figure}

In Figures \ref{fig:2} and \ref{fig:3} we can see the effect of different confinement dimensions. Any potential curve in these calculation
ends for nuclear distance $d=2r_0$, since it was not considered realistic to
further separate away the nuclei while keeping the electron wavefunction
compressed, although this calculation is formally possible.

These curves have been normalized in order to
represent the energy of the process $H^+ + H \rightarrow H_2^+(conf)$, where the
atoms on the left are free. A comparison of the lowest curves in Figure \ref{fig:2}
(for $r_0=\infty$ and $r_0=4$) shows that the minimum is mostly due to a
partial release of the compression energy when the nuclei are displaced
from the barrier surface towards the center, leaving more effective
space to the electron cloud in the Coulomb potential well of the nuclei. This interpretation also provides semiquantitative estimates. For example, the difference between the potential energy surface for $r_0=4$ at the limit value $d=2r_0$ with respect to the corresponding value for $r_0=\infty$ is approximately equal to the energy difference between a $2p$ and $1s$ state of the separated H atom, since the barrier at large $r_0$ for $d=2r_0$ behaves like the corresponding nodal surface.
This interpretation is based on the same concept used in this work to calculate excited states making use of nodal surface, therefore it will be elaborated further. For large enough confinement radius (several atomic units), the effect of the coulomb field generated by the opposite nuclei is no more relevant and, at the same time, the sphere surface in contact with the nuclei can be considered as a plane. This plane coincides with the nodal surface of the corresponding $2p_x$ orbitals in real form \\

$2p_x=\frac{2p_{+1}+2p_{-1}}{i\sqrt{2}} \propto r \exp(-r/2)\cos \theta$. \\

We expect therefore to converge to the corresponding $2p$ orbital energy: this is confirmed by an independent test calculation with a single proton in the origin and a $yz$ nodal plane.
This amount to an energy difference of  $10.2eV$, in agreement with the DMC result. For further illustration, in \ref{fig:1} the potential energy surfaces for the g and u states are reported in a case of mild compression (r$_0$ = 5). It can be seen that both curves reach values close to 10.2 eV at the limit $d = 2r_0$. Calculations demonstrate the convergence of both curves to the same value 10.2 eV for larger confining spheres.
The value of the potential energy surface at the limit abscissa ($2r_0$) increases fast
while increasing compression, due to the effect of pressure on the
electron cloud.

When the radius of the spherical barrier is $\sim 2$ or lower, the
expansion model cannot explain the potential energy surface features. For such strong
compression, an alternative mechanism is based on the idea that, for a
molecule in small containing sphere, the electron wavefunction is
essentially that of the ground state of an electron in a spherical box
$\psi_{01}$. The energy eigenvalue can then be estimated by a
perturbative calculation based on such wavefunction:\\

$E \sim E_{01} - <\frac{2}{|\textbf{r}-{\textbf{i}}\frac{d}{2}|}>_{01} +
\frac{1}{d} + E_H$ \\

The result for $^2\Sigma^+_g$ is shown on Figure \ref{fig:2} and it can be noticed that
the mechanism accounts for the qualitative features of the potential energy surface for low
values of $r_0$ while for a large radius, this approach is less
effective and does not explain the convergence to $10.2 eV$. 

Based on this good agreement, it is possible to use the perturbative result to discuss qualitative features of the potential energy surface. Since the kinetic energy is given by the $E_{01}$ term which is a constant for a giver $r_0$, the attractive part of the potential beyond he minimum is to be attributed to the average Coulomb energy: namely the displacement of nuclei from the barrier surface allows to have higher charge density inside the nuclei potential well. The repulsive part of the potential below the minimum is essentially due to the internuclear repulsion $1/d$.

A very interesting feature emerges from Figure \ref{fig:3}: the $^2\Sigma_u^+$ state,
which is antibonding in the case of the free H$_2^+$ ion, has pronounced
minimum in the case of the compressed ion. These minima have of course
the same explanation as in the case of the enhanced minimum for the
ground state (the previous case). This feature is interesting for
astrophysical radiations transport, since it suggests that the
excitation of $^2\Sigma_g^+$ to the $^2\Sigma_u^+$ state could lead
to radiation diffusion and vibrational excitation of the ground state,
since it cannot result in non radiative dissociation like in the case of
the free ion.

For appropriate use of these results, an estimate of the corresponding pressure of the ground state is provided. This last can be calculated exactly using the relation $p=-(\partial E/\partial V)_T$,  where $E$ is the average energy per molecule and $V$ is the volume of the confining sphere. For an estimate at $T=0$, only the minimum of the calculated potential energy and the zero-point vibrational energy under the harmonic approximation are included. The derivative is also approximated as a finite ratio between two close curves. Under these assumptions, the zero-T pressure is $\sim 9 \times 10^5 atm$ for mild compression $r_0 \sim 2.5$ and raises to $\sim 4.8 \times 10^7 atm$ for more extreme compression $r_0 \sim 1.25$

\section{Conclusions}
\label{concl}
This paper shows that the
application
of DMC method to this problem allows to obtain the energy for any
position of nuclei, confinment shape, using cartesian coordinates. Energy - internuclear distance curves for the ground state ($^2\Sigma_g^+$) and the first excited state ($^2\Sigma_u^+$) of the H$_2^+$ ion under spherical confinement have been calculated for different values of the confinement radius. An appropriate nodal surface is used to select the required excited state. It is found that this molecular ion keeps a classical curve with a neat potential minimum even under conditions of very strong confinement and in spite of the very high energy of the minimum with respect to dissociation to unconfined atoms. Semiquantitative interpretation of the features of the potential energy surface are provided based on two different models for mild and strong confinement. These results support the idea that  H$_2^+$ may be a significant constituent in hydrogen phases under very high pressure astrophysical plasmas. Furthermore, the plots reported may be used as reference in next future studies in view of the lack of alternative, easily accessible data. The technique used is versatile and may be useful to discuss similar systems even under different confinement shapes.

\section*{Acknowledgment}
This research activity has been supported by the General Studies Programme of 
the European Space Agency through contract 4200021790 CCN2.

\bibliographystyle{iopart-num} 
\bibliography{biblio}

\end{document}